\documentclass[aps,prc,twocolumn,superscriptaddress,floatfix,showpacs,lengthcheck]{revtex4}
\usepackage{graphicx}

\begin{document}
\newcommand{\av}{{\rm{av}}}
\newcommand{\fl}{{\rm{fl}}}
\newcommand{\csb}{{\rm{SB}}}
\newcommand{\da}{\downarrow}
\newcommand{\ov}{\overline}
\newcommand{\Ga}{\Gamma}
\newcommand{\ga}{\gamma}
\newcommand{\f}{\frac}  
\newcommand{\be}{\begin{equation}}
\newcommand{\ee}{\end{equation}}
\title{How large is the spreading width of a superdeformed band?}
\author{A.~N.~Wilson}
\email{Anna.Wilson@anu.edu.au}
\affiliation{Department of Nuclear Physics, 
Research School of Physical Sciences and Engineering, 
Australian National University, Canberra, ACT 0200 Australia}
\affiliation{Department of Physics and Theoretical Physics, Faculty of Science, Australian National University, Canberra, ACT 0200 Australia}
\author{A.~J.~Sargeant}
\affiliation{Instituto de F\'isica, Universidade de S{\~{a}}o Paulo,
Caixa Postal 66318, 05315-970  S{\~{a}}o Paulo, SP, Brazil}
\author{P.~M.~Davidson}
\affiliation{Department of Nuclear Physics, 
Research School of Physical Sciences and Engineering, 
Australian National University, Canberra, ACT 0200 Australia}
\author{M.~S.~Hussein}
\affiliation{Instituto de F\'isica, Universidade de S{\~{a}}o Paulo,
Caixa Postal 66318, 05315-970  S{\~{a}}o Paulo, SP, Brazil}
\date{\today}

\begin{abstract}
Recent  models of the decay out of superdeformed bands can broadly be 
divided into two categories.  One approach is based on the similarity 
between the tunneling 
process involved in the decay and that involved in the fusion of 
heavy ions, and builds on the formalism of nuclear reaction theory.  
The other arises from an analogy between the superdeformed 
decay and transport between 
coupled quantum dots.  These models suggest conflicting values for the
spreading width of the decaying superdeformed states.  
In this paper, the decay of superdeformed bands in the five even-even 
nuclei in which the SD excitation energies have been determined experimentally
is considered in the framework of both approaches, and the 
significance of the difference in the resulting spreading widths 
is considered.  The results of the two models are also compared to tunneling
widths estimated from previous barrier height predictions and a 
parabolic approximation to the barrier shape.
\end{abstract}


\maketitle
\section{Introduction}

Superdeformed (SD) nuclei are associated with 
a second minimum in the nuclear potential occurring at large deformation.  The
excited
superdeformed well is separated from the primary minimum
(associated with ``normal'' nuclear shapes) by a real potential barrier.
Rotational SD bands have now been observed in several groups of nuclei with
masses ranging from $A\approx 20$
to $A\approx 240$ \cite{compilation}.  
Although each region displays distinct characteristics
which depend on the underlying nuclear structure supporting the large
deformations, there are some features which are common to all SD
bands.  Perhaps the most interesting of these is the
abrupt decay out of the SD bands to levels of normal
deformation: complete loss of intensity
usually occurs over only two or three consecutive levels.  

\begin{figure}
\includegraphics[width=5.5cm,angle=270]{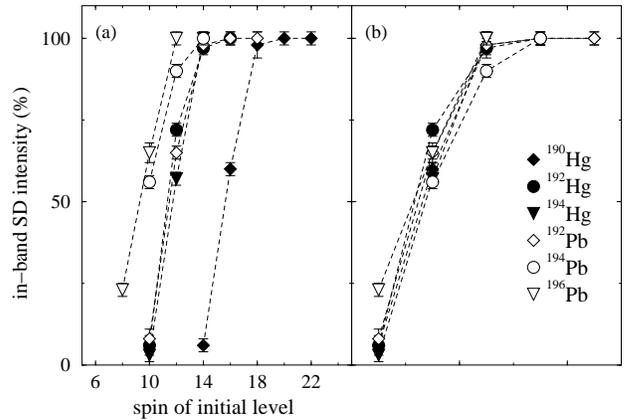}
\caption{Decay profiles of the yrast superdeformed bands 
in $^{190}$Hg, $^{192}$Hg, $^{194}$Hg, $^{192}$Pb, $^{194}$Pb 
and $^{196}$Pb.  The right hand panel shows the profiles shifted in spin
so that the last point represents the lowest spin level from below which some
intensity remains in the band, regardless of the absolute spin.}
\label{fig:intprof}
\end{figure}

In this paper, we focus on the decay of SD bands with masses 
$A\approx 190$ and $A\approx 150$, which are
considered to be two of the classic ``islands of superdeformation.'' 
In these two regions, the extreme deformation is not driven by a small
number of specific single-particle orbitals, but is instead the result
of the complex interplay of macroscopic (Coulomb and rotational) and 
microscopic (shell structure) effects.  Theoretical calculations
indicate that two distinct minima associated with SD and normal
nuclear shapes are present at the point of decay in these nuclei,
unlike, for example, the triaxial superdeformed bands observed in the
$A\approx 160$ region.  Indeed, in the $A\approx 190$ region, the 
SD minimum is thought to exist even at zero rotational frequency. 
Thus it is in these cases that it is clearest 
that the decay occurs via a barrier penetration process.
The intensity profiles of six superdeformed bands in isotopes of
Hg and Pb are shown in Fig.~\ref{fig:intprof}; the data
have been taken from Refs \cite{int190hg}-\cite{intpb196}.
The decay patterns are all remarkably similar, despite the different
excitation
energies and spins of the levels from which the decay occurs.

The possible causes of this rapid loss of flux have been the subject of much
discussion, but despite more than a decade of theoretical and
experimental work, no complete theory of the decay mechanism has been arrived
at.  Indeed, it is as yet not clear whether the sudden enhancement of the
decay probability is due to the dependence on angular momentum of the 
height of the barrier separating the SD and normal potential 
wells, to the increasing effect of pairing correlations 
with decreasing spin, or
to the onset of chaos in the structure of the normal-deformed (ND) states
\cite{hussein}.
One of the obstacles to progress in the attempt to understand the decay
mechanism has been the difficulty in arriving at a consistent, 
broadly-applicable and reliable means of characterizing the size of
the interaction between the SD and ND states
involved in the decay-out.


Recently, two conflicting approaches to the problem
have been proposed which are derived from different 
assumptions concerning the mixing of SD and ND states 
(statistical and two-level mixing).
The conflict manifests itself in the {\it spreading widths} evaluated
for the decaying SD levels, which differ by several
orders of magnitude depending on which model is applied.
It has been suggested \cite{reiner}
that these differences are not important, as
the interaction strengths obtained in analyses of real data are similar
\cite{reiner,mejapan03}.
However, there are several reasons to give the issue serious consideration.

Firstly, the spreading width associated with a particular process is related 
to the strength of the interaction involved.  In the case of the decay
out of the superdeformed well, the relevant interactions are the strong force
(in the rearrangement of the nucleons) and the electromagnetic
force (in the emission of the $\gamma$-ray from the decaying SD level to
the lower-lying ND level).  The spreading widths arrived
at via the two models of the decay differ by orders of magnitude.  
Equally importantly, the parameters on
which the spreading widths depend strongly are different within the two 
approaches, further indicating their incompatibility.

Secondly, in the case of SD nuclei, the spreading width may be
related to the tunneling rate and thus the size of the potential 
barrier separating the superdeformed and normal minima, which should
be comparable with theoretical predictions.  

Thirdly, we want to determine which of the two models,
which have different physical bases and purport to describe the same 
phenomenon, is a more appropriate description of the SD decay.
The concept of a spreading width has proved extremely useful in 
understanding other
issues, such as parity violation, giant multipole resonances,
isobaric analogue states and compound nucleus reactions, and so may help
to distinguish between the two types of model discussed here.

It is the aim of the following
to apply these models to decaying SD levels in nuclei 
with $A\approx 190$ and $A\approx 150$ and to consider the implications 
of the results.  We apply two versions of a statistical mixing model and 
one of a simpler two-level mixing model to all of the yrast SD bands in
even-even nuclei for which excitations and spins have been established.  
This is the first time that all the available data has been treated 
simultaneously, and is also the first time that one of the two statistical
models has been applied to real data.

\section{Descriptions of the decay from the superdeformed well}

Before examining the decay profiles of the bands,
it is useful to summarize the features shared by each of the models
and to look at the differences in the structures of the two classes of model.

\subsection{General assumptions}

There are several basic assumptions common to all treatments of the
SD decay-out process:

(i) The potential barrier separating the SD and ND 
minima is still present at the point of decay;

(ii) The decaying SD states are highly-excited relative to the
yrast (lowest energy for a given spin) ND states;

(iii) The density of ND states
at the same excitation energy and spin as the decaying SD states is high;

(iv) The decaying SD state couples to one or more of these
excited ND states, thus allowing decay to lower-lying states
in the primary minimum.

\begin{figure}
\includegraphics[width=8cm]{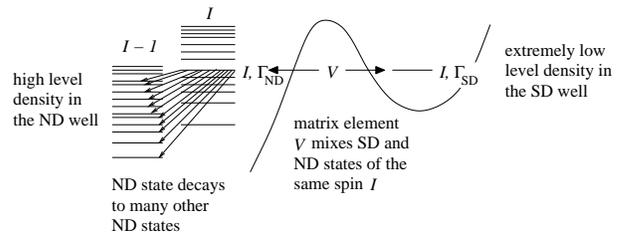}
\caption{Schematic illustration of the SD and ND levels at the point 
of decay.}
\label{fig:sddecay}
\end{figure}

The decay is illustrated schematically in Fig.~\ref{fig:sddecay}.

In general, $I$, 
the fractional intensity remaining in the SD band below
a certain level, is described as depending on four parameters:
$\Gamma_S$, the width for $\gamma$-decay within the 
SD band; $\Gamma_N$, 
the width for $\gamma$-decay of states in the ND 
well; $D$, the average level spacing in the ND well; and $\Gamma^{\downarrow}$,
the spreading width of the decaying SD state.  This last quantity 
measures the fraction of the SD wavefunction extending to the ND well,
and should reflect the size of the barrier separating the two minima.

\subsection{Statistical model}

Recently, a framework originally developed for the study of compound nuclear 
reactions has been used to derive the in- and out-of-band SD intensities.
This approach has developed along two strands: 
(i) the ensemble-averaging
technique of Gu and Weidenm\"uller \cite{gw} (GW), and
(ii) the energy-averaging technique of Sargeant, Hussein and collaborators 
\cite{sarge} (SH). By the ergodic theorem, one expects 
the two averaging techniques to be equivalent.  However, 
although the models are conceptually equivalent, differences in
the derivations - for example, the SH derivation
is strictly valid only in the overlapping resonance regime - mean that
their results will differ in physically realistic cases.  For this 
reason, it is interesting to see how rapidly the SH approach 
deviates from the more widely-applicable GW approach.  For details of the 
differences between the formulations see Refs \cite{gw,sarge}.

\begin{figure}
\includegraphics[width=8.0cm]{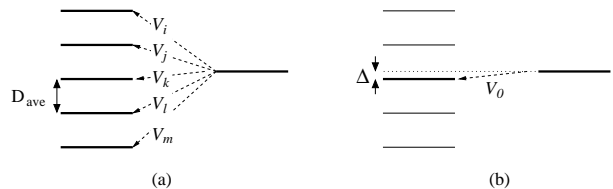}
\caption{(a) In the statistical models, the SD level mixes with many
ND levels of average separation $D$.  
The resulting interaction $\langle V\rangle$ is an average over all of
these interactions.  (b) In the two-level mixing model, the SD level mixes
predominantly with one ND level.  The resulting interaction $V$ is extracted
assuming an average separation of the SD and ND states by 
energy $\Delta = D/4$.}
\label{fig:sdmix}
\end{figure}

In these statistical mixing models, the ND states are assumed to be 
compound, highly-mixed states
which can essentially be treated as ``structure-free''.
The mixing between an SD state and many equivalent energy ND states is 
described statistically; an SD state is assumed to couple
with equal strength to all ND states with the same spin, since their compound
nature implies that there will be no states with wavefunctions which have 
a significant overlap with the wavefunction of the pure SD states.  
The ND states are described by the Gaussian orthogonal 
ensemble (GOE) and the decomposition  
\begin{equation}
I=I_\av+I_\fl
\end{equation}
is made,
where $I_\av$ is an average component
and $I_\fl$  a fluctuating part.
This is in direct analogy with compound nucleus reactions, where 
the cross-section is described by average and
fluctuating (Hauser-Feshbach) components. 
The mixing described by these statistical models is illustrated
schematically in Fig.~\ref{fig:sdmix}(a).

Both GW and SH yield  
\begin{equation}
I_\av = (1 + \Gamma^{\downarrow}/\Gamma_S)^{-1},
\label{eqn:Iave}
\end{equation}
which depends only on the properties of the decaying 
SD state.
As is the case in other fields where a statistical approach is appropriate
(such as reaction theory and atomic physics),
the spreading width is defined by
\begin{equation}
\Gamma^{\downarrow} = 2\pi \langle V\rangle ^2/D,
\label{eqn:spreadV1}
\end{equation}
where $\langle V \rangle$ is the mean interaction 
matrix element of the SD state under study 
and the many nearby ND states.

The fluctuating part $I_\fl$ depends additionally on the properties of the
ND states (the average level separation $D$ and the
$\gamma$-decay width $\Gamma_N$).
The GW approach, which uses supersymmetry techniques to make 
a precise ensemble averaging, results in a non-analytic expression for
the fluctuating part of the intensity.  A fit to the results of
numerical evaluations led to the expression 
\begin{eqnarray}
I_\fl^{\rm{GW}} & =  \left[ 1-0.9139 (\Gamma_N/D)^{0.2172}\right] 
\times \nonumber \\
& \exp \left( \frac{-[0.4343\ln {\Gamma^{\downarrow}/\Gamma_S} - 0.45 
(\Gamma_N/D)^{-0.1303}]^2}{(\Gamma_N/D)^{-0.1477}} \right)
\label{eqn:IflucGW}
\end{eqnarray}
The exact (non-analytic) result of GW is valid for all values
of $\Gamma_N / D$ (that is, not only for systems where the ND states 
approach the overlapping resonance region but also for systems where the
ND states are well-separated and narrow).  Eqn.~\ref{eqn:IflucGW} may not be
valid for all ND conditions, but is expected to be accurate within $20\%$
for $\Gamma_N / D \approx 10^{-4}$ \cite{reiner}, which is close to the 
conditions in some real SD decays.

The SH approach yields the analytic expression
\begin{equation}
I_\fl^{\rm{SH}} 
= 2D/(\pi\Gamma_N) I_\av (1-I_\av)^2,
\label{eqn:IflucSH}
\end{equation}
with $I_\av$ given by Eq.~(\ref{eqn:Iave}).
This expression is strictly valid only in the limit of 
strongly overlapping compound resonances but agrees well the GW expression
down to $\Gamma_N/D $$\sim$$0.1$.  Below this, the SH model may not provide a 
good approximation to the decay.

For all experimentally determined cases,
the fluctuating part of the
intensity is found to be the dominant contribution.

SH also derived the expression  
\be\label{var}
\ov{\left(\Delta I\right)^2}=
{\ov{I_\fl}}^2f_1\left(\xi\right)
+2I_\av\ov{I_\fl}f_2\left(\xi\right),
\ee 
for the mean variance of $I$, which is an energy-integrated version of the 
Ericson fluctuation two-point intensity 
correlation function.  In principle, Eq.~(\ref{var})
permits an assessment of the accuracy with which $\Gamma^{\downarrow}$
and $V$ can be extracted for specific values of $\Gamma_S$,
$\Gamma_N$, $I$ and $D$. The functions $f_1$ and $f_2$ are simple rational
functions of $\xi$=$\Ga_N/(\Ga_S+\Ga^\da)$.

\subsection{Two-level mixing model}

An alternative approach, initially presented by Stafford and 
Barrett \cite{sb} (SB), arose from an analogy between the 
SD decay and transport between coupled quantum dots.
This is essentially a two-state mixing model, in which
the coupling between the SD and ND states is described 
in a Green's function formalism.
The in-band intensity is described by a single component; in this case,
the model describes the situation illustrated in Fig.~\ref{fig:sdmix}(b).
The expression for the in-band intensity $I$ given by Cardamone, 
Stafford and Barrett \cite{csb} can be rearranged to obtain
\begin{equation}
\label{eqn:ISB2}
I=(1+\Ga^\da_\csb/\Ga_S\f{\Ga_N}{\Ga^\da_\csb+\Ga_N})^{-1}.
\end{equation}
The authors of the SB approach state that the correct 
expression for the spreading width from applying Fermi's Golden Rule in 
this model gives a tunneling rate
\begin{equation}
\Gamma^\da_\csb = V^2\f{\Ga_S+\Ga_N}
{\Delta^2 + (\Ga_S+\Ga_N)^2/4}
\label{eqn:fermicsb}
\end{equation}
where $\Delta$ is the energy difference between the interacting ND and 
SD states. 
Here, $V$ is the 
interaction energy of an SD state with a single ND state 
(as opposed to a mean value defined by averaging over the interaction 
of an SD state with many ND states).  When $V$ is extracted from
the data, the unknown quantity $\Delta$ is replaced by an 
average SD-ND level separation given by $|\Delta| = D/4$.
Thus not only are the spreading widths derived in the two models quite 
different quantities, as is clear from their different definitions, 
but so too are the interaction energies $\langle V\rangle$ and $V$.

\subsection{Which class of model is a more appropriate description of 
real superdeformed bands?}

Whilst the above models have been developed to represent the same 
phenomenon,  it can be seen that the statistical and two-level mixing models 
describe different physical situations and therefore
should not necessarily be expected to produce
reconcilable results.  In fact their
application to real data yields very different 
values for the spreading widths $\Ga^\da$ and $\Ga^\da_\csb$.
This raises the question of which is a spreading width in the usual sense.
There is also an important broader
question regarding which {\it type} 
of model describes the 
decay of real SD bands.  In all physically realized situations, the
decay occurs at excitation energies significantly lower than the
overlapping resonance region for the ND levels.  One might thus expect 
the SD state to mix predominantly with only a few ND levels.  In such 
circumstances, does a two-state mixing model provide a better description
than a fully statistical model?

An understanding of which type of model is correct is necessary
before a deeper analysis of the implications of existing data can be made.
Given the present dearth of experimental data establishing excitation
energies and spins of SD levels, one of the few experimental features which 
might be used to interrogate the models is the similarity of the SD bands'
intensity profiles.
The data presented in Fig.~\ref{fig:intprof} show the decay
profiles of bands occurring at quite different excitation energies above
yrast: for example, the decay out of the SD well commences at spin 
$I\approx14\hbar$ and $\approx 3.3$ MeV above 
yrast in $^{192}$Hg \cite{torbenHg},
at $I\approx14\hbar$ and $\approx 4.2$ MeV \cite{khoo,hackman} 
above yrast in $^{194}$Hg,
$I\approx14\hbar$ and $\approx 1.8$ MeV above yrast in $^{192}$Pb \cite{meprl},
and at $I\approx12\hbar$ and $\approx 2.5$ MeV above yrast 
in $^{194}$Pb \cite{waely,boz}.
These excitation energies should correspond to quite
different level densities in the ND well, with the ND level
density for $^{194}$Hg expected to be one to two orders of magnitude higher
than for $^{192}$Pb.  Yet the decay profiles of $^{192}$Pb
and $^{194}$Hg are almost identical, with similar
amounts of intensity leaving the band 
at levels of the same spin in these two SD nuclei.  Any model of the
decay of SD bands should be able to account for this similarity.

If the SD state mixes predominantly with one ND state, as is required
by the two-level mixing model,
such similar decay patterns might not be expected.  
Assuming that the ND levels are complex, ``structure-free'' 
states (as is implicitly done in SB), the strength of
the coupling between an SD and an ND level (and hence the
loss of flux from the SD band at that level) will be governed by the
energy separation $\Delta$ of the two states.  
In some cases, $\Delta$ might approach zero and thus the 
probability for decay out of the band would be very high; in others, 
$\Delta$ might be large and thus the probability for decay out of the 
band would be small.  It should be remembered that SD bands are in real nuclei,
with ND levels at fixed excitation energies.  The assumption
in SB of $\Delta = D/4$ masks the possible effects of the actual distribution
of ND and SD states and their real difference in energy.
It seems surprising that the various values of $\Delta$ for subsequent
levels in all SD bands should be such that the decay profiles
appear the same.  If, on the other hand, the ND states are not complex,
the coupling between the SD and ND levels should be affected by the 
underlying microscopic structure of the ND level.  Such circumstances
also seem unlikely to produce near-identical decay profiles in different
nuclei.

\begin{figure}
\includegraphics[width=8cm]{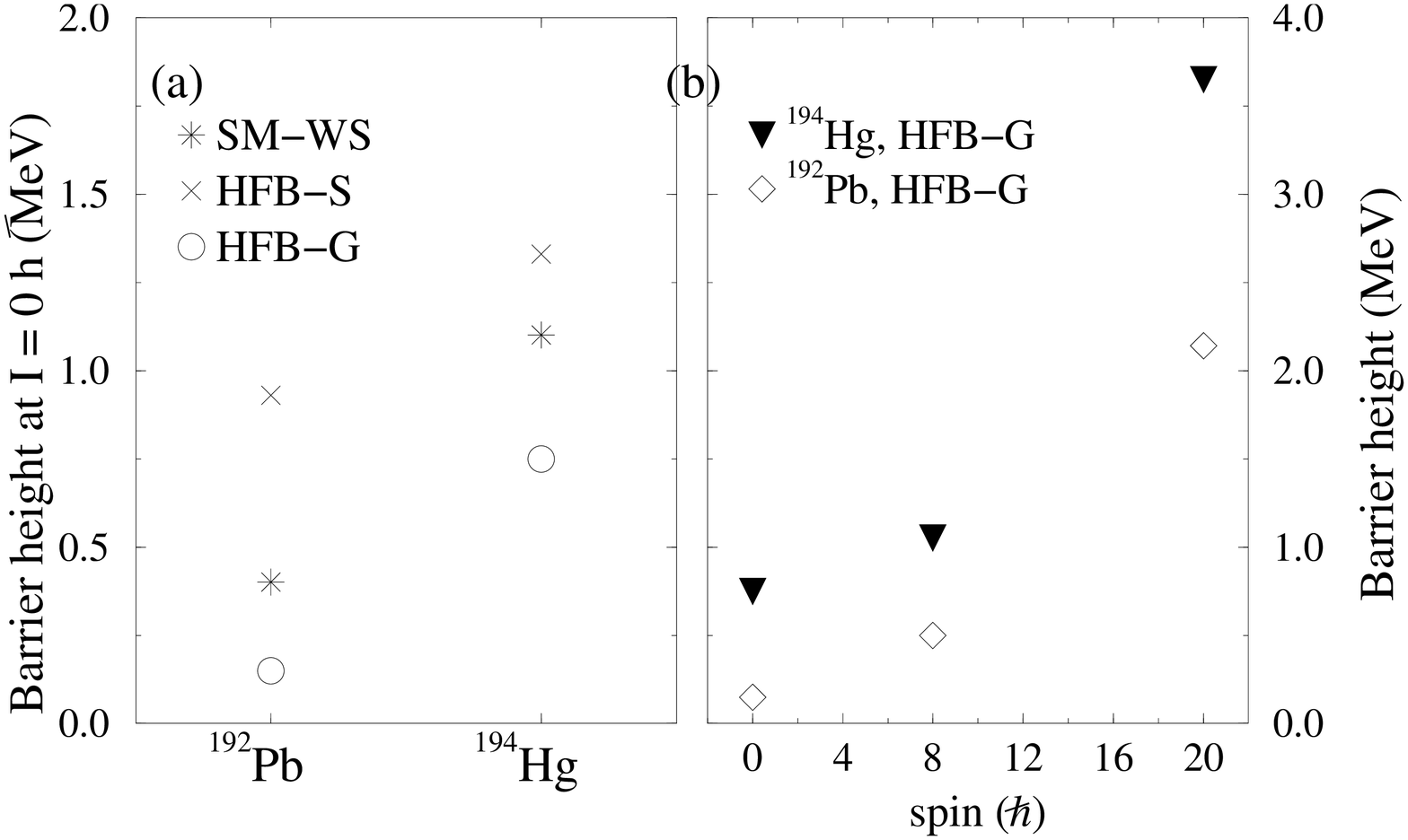}
\caption{(a) Comparison of various model predictions of the height of the 
barrier separating the SD and ND wells in $^{192}$Pb and $^{194}$Hg: Strutinsky
method with Woods-Saxon potential (SM-WS) \cite{satula},
mean field using the Skyrme interaction (HFB-S) \cite{krieger} and 
mean field using a Gogny interaction (HFB-G) \cite{libert}.
(b) Comparison of the barrier heights predicted using the HFB method and a
Gogny force for $^{192}$Pb and $^{194}$Hg at various different spins.} 
\label{fig:barrierpred}
\end{figure}

The alternative model is completely statistical mixing between SD and ND
states, so that each SD 
level mixes with many equivalent ND levels of the same spin/parity.
It should then be expected that an increased
level density in the ND well would
result in an increased probability for decay out of the SD minimum from 
a level of fixed $\Ga_S$ and $\Ga^\da$.  
In this  scenario, the lower level
density in $^{192}$Pb must be compensated for by a 
correspondingly lower barrier height (and hence larger $\Ga^\da$) 
if the similarity of the decay profiles is to be accounted for.
Predictions of the barrier heights
in $^{192}$Pb and $^{194}$Hg at spin $I=0\hbar$
have been made using Strutinsky \cite{satula}
and Hartree-Fock-Bogoliubov (HFB) \cite{krieger} methods;
more recent HFB calculations \cite{libert} have predicted
barrier heights for these two nuclei
at spins $I=0\hbar$, $I=8\hbar$ and $I=20\hbar$.
The results of these calculations are shown in 
Fig.~\ref{fig:barrierpred}.  All 
three methods predict that the barrier height for $^{194}$Hg is larger
than for $^{192}$Pb at spin zero, and the predictions at non-zero spin 
suggest that the difference remains large and indeed increases towards 
high spins.

Information concerning relative barrier heights
and the stability of the SD well in different nuclei is
one of the primary goals of studies of superdeformation.  Without
knowing whether the mixing is completely statistical or predominantly
with one ND level, or the degree to which it is affected by the
microscopic structure of the ND states, quantitative comparisons
between data and model predictions cannot be made.  
The problem is analogous to that of pinpointing the reaction
mechanism in scattering.  If  an oversimplified
or inappropriate reaction mechanism is assumed, the nuclear structure information
obtained will be inconclusive.
\section{Application of the models to the existing data: spreading width results}

Before the three models of the decay described in the previous section 
can be applied to analyse SD decay profiles, 
values of $D$, $\Gamma_N$, $\Gamma_S$ and $I$
are needed.  In the following, we briefly describe how the values used in this
work have been arrived at.

Estimates of $D$ are usually made using a Fermi gas density
of states. 
The decay widths of the ND states, $\Ga_N$, can be estimated using the
Fermi gas model of the level density and 
the additional assumption that the compound
ND states decay predominantly by statistical E1 emission.  
To arrive at these estimates, the excitation energies of the SD states
to be known.
To date, unambiguous determination of
excitation energies and spins
has only been possible in
five nuclei in the $A$$\approx$$150$ and $A$$\approx$$190$
regions:  $^{194}$Hg \cite{khoo,hackman},
$^{194}$Pb \cite{waely,boz}, $^{152}$Dy \cite{lauritsen},
$^{192}$Pb \cite{meprl} and $^{191}$Hg \cite{siem}.  
The method of analysing the quasicontinuum component
of the SD decay spectrum has provided a somewhat less precise
measurement in $^{192}$Hg \cite{torbenHg}.  
In this paper, we consider all five $K=0$ ``vacuum'' bands in the 
even-even isotopes.

The estimates of $D$ and $\Ga_N$ also rely on 
an assumed ``backshift parameter'', 
which corresponds to the
gap in the level density above the yrast line in even-even nuclei due to
pairing correlations.
In the following, the usual value of the backshift parameter ($1.4$ MeV) is
adopted in the treatment of $^{192,4}$Hg, while
the spin-dependent 
backshift parameters suggested 
in Ref \cite{meandpaul} are adopted in the treatment of
$^{192,4}$Pb.  However, it should be noted that these values
have uncertainties such that they may increase or decrease by a factor
of two or more \cite{mejapan03,meandpaul}.

Values of $\Gamma_S$ can be extracted from the data with relatively
small uncertainty (of the order of 10$\%$) if lifetime measurements have
been made and quadrupole moments extracted: such measurements have been
made for all of the nuclei considered here \cite{busse}-\cite{medsam}.
The final parameter needed to allow a spreading width to be extracted is 
$I$, the fraction of intensity that remains in the SD band.  In
general, with the high statistics experiments which have been performed
in recent years, this is measured to an accuracy of $\sim 2\%$ 
of the maximum in-band intensity.  The 
values adopted here are taken from Refs \cite{reiner,meprl,waely}.
Table~\ref{tab:inputs} lists $D$, $\Gamma_N$,
$\Gamma_S$ and $I$ for two levels from which significant 
flux is lost in the yrast
SD bands in $^{192,194}$Hg and $^{192,194}$Pb.

\begin{table}
\caption{\label{tab:inputs}Values of ND level densities, ND 
and SD $\gamma$-decay widths and in-band intensity fractions 
for two levels in the yrast SD bands in $^{192,194}$Hg 
and $^{192,194}$Pb.}
\begin{ruledtabular}
\begin{tabular}{cccccc}
Isotope & Spin $(\hbar)$ & $D$ (eV) & $\Gamma_N$ ($\mu$eV) & $\Gamma_S$ ($\mu$eV)& $I$\\
\hline
$^{192}$Hg & $10$ & $89$  & $733$  & $ 50$ & $0.08$\\
$^{192}$Hg & $12$ & $135$ & $613$  & $128$ & $0.74$\\
$^{194}$Hg & $10$ & $14$  & $1487$ & $33 $ & $\leq0.05$\\
$^{194}$Hg & $12$ & $19$  & $1345$ & $86 $ & $0.60$\\
\hline
$^{192}$Pb & $ 8$ & $1681$& $169$ & $ 16 $ & $\leq0.25$\\
$^{192}$Pb & $10$ & $1410$& $188$ & $ 48 $ & $0.12$\\
$^{194}$Pb & $ 6$ & $333$ & $405$ & $ 3 $ & $\leq0.04$\\
$^{194}$Pb & $ 8$ & $273$ & $445$ & $ 14$ & $0.65$\\
\end{tabular}
\end{ruledtabular}
\end{table}

Spreading widths extracted using 
the GW, SH and SB approaches, are listed in table~\ref{tab:spreads}.
Equivalent calculations for the spin $26\hbar$ and spin $28\hbar$ levels 
in $^{152}$Dy are included in the table: 
the values of $D$, $\Gamma_N$, $\Gamma_S$ 
and $I$ for these states are those given in Ref~\cite{lauritsen}.
The standard deviations in the in-band SD intensity
calculated within the SH model
are included in the right-most column.
The small values obtained indicate that, within its range of
validity, the model predicts well-defined values of $I$.

\begin{table}
\caption{\label{tab:spreads}
Spreading widths of SD levels in
$^{192,4}$Hg, $^{192,4}$Pb and $^{152}$Dy
calculated using the GW, SH and SB models.  
The final column gives the standard deviation calculated using 
the SH approach.}
\begin{ruledtabular}
\begin{tabular}{ccccc}
&\multicolumn{2}{c}{$\Gamma^{\downarrow}$ (eV)} & $\Gamma^{\downarrow}_{\rm{SB}}$ (eV)  & $(\overline{(\Delta I)^2})^{1/2}$ \\
&GW&SH&SB&SH\\
\hline
$^{192}$Hg($10$)         & $30.5$          & $48.0 $         & $2.7\times 10^{-3}$     & $5\times10^{-4}$\\
$^{192}$Hg($12$) 	 & $0.35$          & $24.1 $         & $4.9\times 10^{-5}$     & $6\times10^{-3}$\\
$^{194}$Hg($10$)	 & $\geq 2.12$	   & $\geq 3.96$     & $\geq 1.1\times 10^{-3}$& $1.5\times10^{-3}$ \\
$^{194}$Hg($12$)	 & $0.058$ 	   & $1.29$ 	     & $6.0\times10^{-5}$      & $3.0\times10^{-2}$ \\
\hline
$^{192}$Pb($8$) 	 & $\geq 510$ 	   & $\geq 405 $     & $\geq 6.7\times 10^{-5}$& $2.5\times10^{-4}$ \\
$^{192}$Pb($10$)         & $7509$          & $1910$          & $-4.0\times 10^{-3}$    & $5.9\times10^{-5}$ \\
$^{194}$Pb($6$)          & $\geq 99.9$     & $\geq 39.3 $    & $\geq 8.8\times10^{-5}$ & $2.0\times10^{-4}$\\
$^{194}$Pb($8$)          & $0.223$         & $8.41$          & $7.7\times 10^{-6}$     & $7.5\times10^{-3}$ \\
\hline
$^{152}$Dy($26$) 	 & $68$            & $267 $ 	     & $-0.04$ 	        & $2.4\times10^{-3}$\\
$^{152}$Dy($28$) 	 & $6$ 	           & $137$	     & $ 0.01$ 	        & $1.06\times10^{-2}$\\
\end{tabular}
\end{ruledtabular}
\end{table}


Before discussing the results in detail, we should briefly comment on the 
fact that, for some levels,
the value extracted for $\Gamma^{\downarrow}_{\rm{SB}}$ 
is negative.
As previously noted \cite{csb,meandpaul}, the necessity that
the spreading width is positive imposes the additional
constraint that  $\Gamma_{N} > \Gamma_{S} (I^{-1} - 1)$ in this model.
In order to obtain a positive
value for $\Gamma^{\downarrow}_{\rm{SB}}$ for the anomalous 
states in $^{192}$Pb and $^{152}$Dy, the ND $\gamma$-decay
width would need to be increased by only a factor of $2-3$. 
Given the uncertainties and assumptions employed in
estimating $D$ and $\Gamma_N$, such an increase is not inconceivable.
The negative values should therefore not be taken as an indication of 
the failure of the model.

\section{Comparison of the spreading widths and discussion}

\begin{figure}
\includegraphics[width=5.4cm,angle=270]{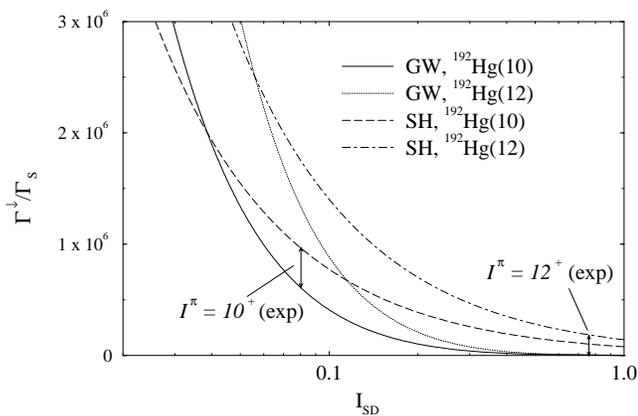}
\caption{Values of the ratio $\Gamma^{\downarrow}/\Gamma_S$ as a function
of the in-band intensity $I$ calculated for
$\Gamma_N/D$ appropriate for the $10^+$ and $12^+$ levels
in $^{192}$Hg.}
\label{fig:gwsh}
\end{figure}

Fig.~\ref{fig:gwsh} compares the models of GW and SH
for $\Gamma_N/D$ equal to the estimated ratios for the
two SD levels in $^{192}$Hg.  The curves illustrate the dependence
of the calculated ratio $\Gamma^{\downarrow}/\Gamma_S$ as a function
of $I$.  For in-band intensities in the range $0.03 - 0.1$,
the values obtained with SH are within an order of magnitude of
those obtained with GW.  However, for higher
intensities, the values obtained with GW drop much more rapidly
and the difference between the two approaches becomes several
orders of magnitude.
Thus although GW and SH are essentially equivalent in the overlapping
resonance region, SH deviates from GW for $\Gamma_N/D$$\ll$$1$ and
$I$$\ge$$0.1$.

It has been suggested that the GW
fit formula is accurate to within about $20\%$ for nuclei with 
$A$$\approx$$190$ \cite{reiner}; SH is valid over a significantly smaller 
range of $\Gamma_N/D$.  In the cases considered here (which 
are typical of superdeformed bands in the $A\approx 190$ 
and $A\approx 150$ regions), $\Gamma_N/D$$\ll$$1$ and thus 
SH should not be expected to provide a precise model. 
However, in most cases, 
the values of $\Gamma^{\downarrow}$ are comparable (within one or
two orders of magnitude)
between the two models.  They thus
yield comparable values for the mean interaction
strength, since both models relate  $\Gamma^{\downarrow}$ 
and $\langle V \rangle$ through Eqn.~\ref{eqn:spreadV1}.

The values of $\Gamma^{\downarrow}_{\rm SB}$, on the other hand, are several
orders of magnitude smaller than $\Gamma^{\downarrow}$
in all cases.
Similar differences were found in Refs.~\cite{csb,reiner}.
For the Hg and Pb isotopes,
$\Gamma^{\downarrow}_{\rm{SB}}/ \Gamma^{\downarrow}$ is in
the range $10^{-3} - 10^{-6}$.
The values of  $\Gamma^{\downarrow}_{\rm{SB}}$
obtained for the two
levels in $^{152}$Dy are also
significantly less than $\Gamma^{\downarrow}$.
As noted above, these differences have previously
been dismissed as unimportant because the interaction energies
extracted in the two types of model have been found to be similar.
In the following, we consider reasons why this attitude may gloss over
basic differences in the models.

Firstly, finding the correct value of the spreading width should help 
understand its physical significance.
The values of $\Ga^\da_\csb$ are very small, and in fact those
extracted for the Pb isotopes approach
the experimental value of the weak interaction spreading width 
$\Ga_{\rm{W}}$=$1.8\times 10^{-7}$ eV obtained in parity violation studies
with epithermal neutrons \cite{Mitchell:2001}. 
The spreading width for decay out of an SD band describes 
a rearrangement of nucleons due to the strong and 
electromagnetic interactions.  It might therefore be expected to be of
similar magnitude to the spreading widths encountered in other
nuclear and electromagnetic processes, {\it i.e.} of the order of MeV and eV
respectively.  The spreading widths obtained with GW are of this
order.  The drastic nature of the nuclear shape change may indeed result
in a smaller spreading width than is encountered in other strong/EM processes,
but the extreme smallness of $\Gamma^{\downarrow}_{\rm SB}$ needs
to be understood if the SB model is put forward
as an appropriate description of the SD decay and 
if $\Gamma^{\downarrow}_{\rm SB}$ is truly a spreading width.

Secondly, there is the issue of the relation between the spreading width and 
tunneling rate, and by extension the relation with the size of
the barrier separating the SD and ND minima.  

If the mixing is statistical and there is a quasicontinuum of ND states, 
it may be possible to relate the height of the barrier to a 
fusion-like tunneling rate.
Assuming that the barrier and the two wells
can be modeled with parabolic and inverse 
parabolic shapes, the barrier height $B$ can be related to a 
tunneling rate $\Gamma_{tunnel}$ 
by the relation \cite{shimizu,reiner}
\begin{equation}
\Gamma_{tunnel} = \frac{\hbar\omega_s}{2\pi}{\rm e}^{(- \frac{2\pi B}{\hbar \omega_b})},
\label{eqn:barrier}
\end{equation}
where $\omega_s$ and $\omega_b$ specify 
the curvatures of the parabolas describing the SD well and the 
barrier respectively. Making the further assumption
that $\hbar \omega_s = \hbar \omega_b = 0.6$ MeV \cite{reiner}, it is possible
to use the barrier heights predicted by the Strutinsky or HFB calculations
to predict tunneling rates.  

As an illustration, we focus on the two nuclei 
with lowest and highest measured SD excitation energies, $^{192}$Pb and
$^{194}$Hg.  In both cases, the decay out of the SD band occurs
in the spin range $I\approx 14\hbar\rightarrow I\approx 8\hbar$.  
Barrier height 
predictions at spins $I=8\hbar$ and $I=20\hbar$ have been made for both of
these nuclei using the HFB method with a Gogny force \cite{libert}.  
The results
of these calculations (shown in Fig.~\ref{fig:barrierpred}(b)) have been used
to estimate tunneling widths as described above.  The resulting tunneling
widths are given in table~\ref{tab:barrierspreads}.

\begin{table}
\caption{\label{tab:barrierspreads}Tunneling widths for escape from the
SD well extracted using Eqn.~\ref{eqn:barrier} and the barrier heights
predicted in Ref. \cite{libert}.}
\begin{ruledtabular}
\begin{tabular}{lll}
Isotope & Spin & $\Gamma_{tunnel}$ (eV) \\
\hline
$^{194}$Hg & $0$ & 37\\
$^{194}$Hg & $8$ & $1.6$\\
$^{194}$Hg & $12$ (interpolated) & $0.38$\\
$^{194}$Hg & $20$ & $2.4\times10^{-12}$\\
\hline
$^{192}$Pb & $0$ & $1.98\times10^{4}$\\
$^{192}$Pb & $8$ & 508\\
$^{192}$Pb & $10$ (interpolated) & 30 \\
$^{192}$Pb & $20$ & $1.77\times10^{-5}$\\
\end{tabular}
\end{ruledtabular}
\end{table}

The tunneling width associated with the predicted barrier
height at spin $I=8\hbar$ is close to the lower limit of the spreading width 
extracted using the two statistical models, and several orders
of magnitude larger than the lower limits extracted using the
two-level mixing model.  However,  in the limit of 
mixing with only a small number of ND levels the relationship between
barrier height and tunneling width given in Eqn.~\ref{eqn:barrier}
is not appropriate \cite{bjornholmlynn}.

In order to compare the tunneling rate with the spreading
width of GW for levels where definite values of $\Gamma^{\downarrow}$ and
$\Gamma^{\downarrow}_{\rm SB}$ are obtained, we have estimated barrier
heights at $I=12\hbar$ and $I=10\hbar$ in
$^{194}$Hg and  $^{192}$Pb respectively.  Using a simple linear
interpolation between $I=8\hbar$ and $I=20\hbar$, we obtain
$B(I=12\hbar,{\rm ^{194}Hg}) = 1.98$ MeV and 
$B(I=12\hbar,{\rm ^{192}Pb}) = 0.77$ MeV
The tunneling widths corresponding
to these values are included in the table.  
For $^{194}$Hg, $\Gamma^{\downarrow}$ obtained with GW is an order of 
magnitude smaller than the estimated tunneling width 
($\Gamma^{\downarrow}_{\rm SB}$ is four orders of magnitude smaller);
for $^{192}$Pb, it is two orders of magnitude larger (comparison with 
$\Gamma^{\downarrow}_{\rm SB}$, which is negative in this instance, is 
not possible).

It is impossible to evaluate the significance of these differences without
a firm idea of the character of the mixing between the SD and ND states.
If the coupling between SD and ND levels is fully
statistical, $\Gamma^{\downarrow}$ and the expression for
$\Gamma_{tunnel}$ are meaningful.  However, the relatively low ND level density
in the two Pb isotopes (where the overlapping
resonance region and chaos are not approached) indicates that a fully
statistical treatment may not be appropriate.  
If this is the case, the systematics of the SD minima in the $A\approx 190$
region cannot be studied without further theoretical development
of the decay-out models.

Finally, we consider whether the similarity of the
interaction energies obtained in previous work \cite{reiner,mejapan03}
should be considered natural, or even desirable.
As illustrated in Fig.~\ref{fig:sdmix}, the interactions $\langle V\rangle$ 
and $V$ are not the same and there is therefore no {\it a priori} reason
why they should have the same value.  In the GW (and SH)
approach, $\langle V \rangle$ describes the averaged 
interaction of the 
SD state with many ND states.
In contrast, in the SB
approach, $V$ is the interaction between the SD state and 
a single ND state.
Because of these different physical meanings, $\langle V\rangle$ and $V$
depend on different parameters.  While $\langle V\rangle$ 
and  $\Gamma^{\downarrow}$ depend on the ND level spacing $D$,
$V$ (but not  $\Gamma^{\downarrow}_{\rm{SB}}$) depends on $\Delta$, the energy
difference between the SD state and the ND state it mixes with.

For any real decaying SD level $\Delta$ is unknown, and the
resulting interaction strength
depends strongly on the value adopted.
As an example, the GW approach yields $\langle V \rangle \approx 20$ eV 
for the $10^+$ level in $^{192}$Hg: the SB approach yields
$V \approx 0.03$ eV for $\Delta = 0$, but any higher value can be obtained,
including $20$ eV if $\Delta \approx 1.2 D$.
In previous work \cite{reiner,meandpaul}
the average SD-ND energy difference $\Delta = D/4$ has been adopted, 
resulting in $V$ of similar size to $\langle V \rangle$.  However,
there is no obvious reason to expect that
use of the average separation should 
produce the same effect as averaging the coupling to many ND states. 
In effect, the GW/SH models offer a means of experimentally 
determining $\langle V\rangle$ 
and  $\Gamma^{\downarrow}$, whereas 
SB offers a means of experimentally determining $\Gamma^{\downarrow}_{\rm{SB}}$
{\it but not} $V$, unless $\Delta$ can be measured (or calculated).  The
dependence of $I_{\rm av}$ on $\Delta$ is investigated for a related model
in Ref. \cite{sarge2}.

The underlying differences between the two types of model are exemplified
by the different treatments of the
in-band intensity as comprising either one or two contributions.
In the analyses using the GW and SH models, 
$I_\av$ is found to be extremely small and $I_\fl$
dominates.  In the SB model, there is no separate 
fluctuating contribution, and the total intensity is given by
Eq.~(\ref{eqn:ISB2}).  However, in the limit where
$\Gamma_N \gg \Ga^\da_\csb$, 
this reduces to $I\rightarrow(1+\Ga^\da_\csb/\Ga_S)^{-1}$,
which is {\it formally} identical to $I_{\av}$.
The data gathered to date suggest that 
this limit is approached in most physically realized cases.

The motivation behind the SB model may be thought of as 
similar to the calculation of 
a ``doorway-doorway'' interaction in the GW/SH approach, for which
a fluctuation contribution could also be calculated \cite{hussein}.
It seems unlikely, given the relatively low ND level density,
that a fully statistical mixing process occurs in the decay of
SD $^{192}$Pb, and thus this type of doorway approach may be 
more suitable than the present statistical models.  In addition, the idea of 
a doorway state, which would have larger overlap with an SD state than 
other nearby ND states, goes some way towards making allowance for 
the possible influence of the microscopic structure of the ND states 
on the decay.  
If there are indeed ND doorway states then these in turn will have a spreading
width $\Ga^\da_d$ and a $\gamma$-emission (escape) width $\Ga_d$, analogous
to $\Ga^{\downarrow}$ and $\Ga_S$ for the in-band SD state. The ND
compound width, $\Ga_N$, would then be reduced (due to flux conservation)
to an extent which is measured by an additional parameter,
$\mu=\Ga^\da_d/\Ga_d$; a similar picture has been used to describe the
decay of giant resonances \cite{dias,solokovandzelevinsky}. 
Note that the spreading widths $\Ga^\da$ and $\Ga^\da_d$ depend on the 
choice of basis in which
the partial diagonalizations which define them are carried out.
However, without a reliable calculation of $\mu$ its introduction would
not be useful in the context of the present paper: analysis of
the experimental intraband intensity, $I$, at present 
permits the determination of a single parameter, $\Ga^\da$. 
It may be possible to determine $\mu$ from an analysis of the
quasicontinuum component of the decay spectrum \cite{sven2,waelyepj},
but it is likely that experimentally distinguishing
$\Ga_N$ and $\Ga_d$ will be extremely difficult.

\section{Conclusion}

In this paper, we have examined the structure of three recent models 
of SD decay, and applied these
models to the analysis of the decay of 
states in four SD bands in nuclei with masses $A$$\approx$$190$ and 
one SD band in $^{152}$Dy.  These bands are the only yrast SD bands in 
even-even nuclei for which excitation energies have been measured, and
are thus the only such cases where quantitative estimates of the normal
deformed state properties can be made.  This work represents the first time
all of the available experimental data have been compared.
Two of these models are fully statistical,
and describe the physical process where an SD state mixes with many 
ND states that have structures sufficiently complex so as to be equivalent.
The other model assumes that the mixing occurs predominantly with
only one ND state (the nearest neighbor).  Questions as to the nature of
this state, whether complex or of well-defined microscopic origin, are
not addressed by the model.

Since the two types of model represent
physically different processes, they should not be expected
to result in the same interaction energy -- that is, it should not
be expected that $\langle V\rangle$ of the statistical mixing model
be equal to $V$ of the two-level mixing model.  
We have stressed the point made by the authors of the two-level mixing model
\cite{sb,csb}
that $V$ cannot be extracted from the data without knowledge of the SD-ND
energy separation $\Delta$; in relation to this we have questioned whether the
use of the average separation in that model should be expected to
be equivalent to the use of the
average over many interactions in the statistical model.  

Because of the
present impossibility of arriving at a definite value of $V$ for the
decay from any measured SD level,
comparison should be restricted to a directly calculable quantity such as
the spreading width.
The energy-averaging SH approach results in similar
spreading widths to those
extracted in the equivalent ensemble-averaging GW approach, but as expected the
SH model fails more quickly than the GW fit formula 
as $\Gamma_N / D \rightarrow \ll 1$.
The results
of these statistical mixing approaches are consistent with expectations 
for electromagnetic/strong interaction spreading widths.
They are also closer to the range of tunneling widths obtained
with a simple model of the barrier potential and the predictions
of HFB calculations.
The two-level mixing approach results in far smaller spreading widths, such
that some special explanation of their size would be required if this model
of the decay is correct.  

The different definitions of the spreading widths indicate that they are 
incompatible quantities.  But the question of which is ``correct'' is 
intimately related to which type of model is physically more appropriate.  
This question is particularly important given
the similarity of the decay patterns of SD bands over a wide
range of excitation energy above yrast, and hence over a variety of different 
ND level densities.  The experimentally determined SD excitation energies
are also beginning to indicate that, at least in some cases such as the two
Pb isotopes, it is unlikely that the ND states are truly compound.  Thus
it appears that some consideration of the microscopic structure of the ND
states may be required, possibly by the introduction of doorway states.
Until it is established which model describes the real
process of decay out of the SD well, a deeper understanding of what triggers
the decay and why the intensity profiles are so similar cannot be
achieved.

\begin{acknowledgments}
This work was carried out with support from the Australian Research 
Council through grant DP0451780.
MSH is partly supported by FAPESP and the CNPq and both MSH and AJS are 
supported by the Instituto de Mil\^enio de 
Informa\c c\~ao Qu\^antica - MCT, all of Brazil.
\end{acknowledgments}

\bibliographystyle{unsrt}

\end{document}